\documentclass[12pt]{article}
\usepackage{graphicx}
\usepackage{amsmath}
\usepackage{amssymb}
\usepackage{caption2}
\begin{document}
\bibliographystyle{prsty}
\begin{center}
{\large {\bf \sc{   Ground state  mass spectrum for  scalar diquarks   with Bethe-Salpeter equation }}} \\[2mm]
Z. G. Wang$^{1}$ \footnote{Corresponding author; E-mail,wangzgyiti@yahoo.com.cn.  }, S. L. Wan$^{2}$ and W. M. Yang$^{2} $    \\
$^{1}$ Department of Physics, North China Electric Power University, Baoding 071003, P. R. China \\
$^{2}$ Department of Modern Physics, University of Science and Technology of China, Hefei 230026, P. R. China \\
\end{center}

\begin{abstract}
In this article, we study  the structures of the pseudoscalar mesons
$\pi$, $K$ and the scalar diquarks $U^a$, $D^a$, $S^a$  in the
framework of the coupled rainbow Schwinger-Dyson equation and ladder
 Bethe-Salpeter equation with the confining effective potential.
 The $u$, $d$, $s$ quarks have    small current masses,
 and the renormalization is very large, the mass poles in the timelike region are  absent which
  implements  confinement naturally.
 The Bethe-Salpeter wavefunctions of the pseudoscalar
mesons $\pi$, $K$ and the scalar diquarks $U^a$, $D^a$, $S^a$
 have the same type (Gaussian type) momentum dependence,
  center around  zero momentum and extend to the energy scale about
$q^2=1GeV^2$ which happen to be the energy scale for the chiral
symmetry breaking, the strong interactions  in the infrared region
result in bound (or quasi-bound) states. The numerical results for
the masses and decay constants of the  $\pi$, $K$ mesons can
reproduce the experimental values, the ground state masses of the
scalar diquarks $U^a$, $D^a$, $S^a$ are consistent with the existing
theoretical calculations. We suggest a new Lagrangian which may
explain  the uncertainty of the masses of the scalar diquarks .
\end{abstract}


\section{Introduction}
 The discovery  of the pentaquark state $\Theta^+(1540)$ has opened a new
field of strong interaction  and provides a new opportunity for a
deeper understanding of the low energy QCD. Intense theoretical
studies have been motivated  to clarify the quantum numbers and to
understand the under-structures of the pentaquark state
$\Theta^+(1540)$ \cite{ReviewPenta}. Although the existence of the
$\Theta^+(1540)$ is uncertain and is a subject of controversy now,
the $\Theta^+(1540)$ has already contributed to hadron spectroscopy.
Unlike the chiral soliton model, the
 quark models take the constituent quarks or quark clusters as the elementary
degrees of freedom,  there exist a great number of  possible quark
configurations satisfy the Fermi statistics and the color singlet
condition  for the substructures of the pentaquark state
$\Theta^+(1540)$   if we release  stringent dynamical constraints.
 In fact, the multiquark states
are many-body problems, they are very difficult to solve. Whether or
not the quarks can cluster together to form diquarks is of great
importance theoretically, if we take the diquarks as the basic
constituents (here the "basic constituents" does  not mean they are
asymptotic  states, they  just exist inside the baryons or
multiquark states with typical length),   the problems will be
greatly simplified, for example, the $\frac{1}{2}^+$ octet and
$\frac{3}{2}^+$ decuplet baryons can be taken as
 quark-diquark bound states  in the  Faddeev approximation \cite{Faddeev},
 the nonet scalar mesons below $1GeV$ can be taken as 4-quark
states $(qq)_{\bar{3}}(\bar{q}\bar{q})_3$ with scalar diquarks or
pseudoscalar diquarks as their basic constituents \cite{WangScalar};
furthermore, we can obtain more insight into the relevant degrees of
freedom and deepen our understanding about the underlying dynamics
that determines the properties of the baryons and exotic multiquark
states.   A typical quark model for the pentaquark  states is the
Jaffe-Wilczek's diquark-diquark-antiquark model \cite{Jaffe03}. In
this model, the scalar diquarks  $U^a= \epsilon^{abc}
d_b^T(x)C\gamma_5 s_c(x) $, $D^a=\epsilon^{abc} u_b^T(x)C\gamma_5
s_c(x)$ , $S^a=\epsilon^{abc} u_b^T(x)C\gamma_5 d_c(x)$ are taken as
the basic constituents. They belong to the antitriplet $\bar{3}$
representation of both the color $SU(3)_c$ group and flavor
$SU(3)_f$ group, in the color superconductivity theory, the
attractive interactions in this channel lead to the formulation of
nonzero condensates and  breaking of both the color and flavor
$SU(3)$ symmetries \cite{ReviewColor}. The scalar diquarks
correspond to the $^1S_0$ states of the diquark systems,  the
one-gluon exchange force and the instanton induced force can lead to
significant attractions between the quarks in the $0^+$ channels
\cite{GluonInstanton}. The pseudoscalar diquarks do not have
nonrelativistic limit,  can be taken as  the $^3P_0$ states. As the
instanton induced force results in strong attractions in the scalar
diquark channel and strong repulsions in the pseudoscalar diquark
channel, if the effects of the instanton are manifested, we prefer
the $S$ type diquark to the $P$ type diquark in constructing
interpolating currents in the QCD sum rules \cite{GluonInstanton}.

In this article, we take the point of view that the scalar diquarks
are quasi-bound states of quark-quark system and study  the ground
state mass spectrum within the framework of the coupled
Schwinger-Dyson equation (SDE) and Bethe-Salpeter equation (BSE).
   The coupled rainbow SDE and ladder BSE have given
a lot of successful descriptions of the long distance properties of
the low energy QCD and the QCD vacuum (for reviews, one can see
Refs.\cite{Roberts94,Tandy97,Roberts00,Roberts03}). The SDE can
naturally embody the dynamical symmetry breaking and confinement
which are two crucial features of QCD, although they correspond to
two very different energy scales \cite{Miransky93,Alkofer03}. On the
other hand, the BSE is a conventional approach in  dealing with the
two-body relativistic bound state problems \cite{BS51}. From the
solutions of the BSE, we can obtain useful information about the
under-structures of the mesons and  diquarks, and   obtain powerful
tests for the quark theory. However, the obviously drawback may be
the model dependent kernels for the gluon two-point Green's function
and the truncations for the coupled divergent SDE and BSE series in
one or the other ways \cite{WYW03}. Many analytical and numerical
calculations indicate that the coupled rainbow SDE and ladder BSE
with phenomenological potential models can give model independent
results and
 satisfactory values \cite{Dai,MarisExample}. The usually used
effective potential models are confining Dirac $\delta$ function
potential, Gaussian  distribution potential and flat bottom
potential (FBP) \cite{Munczek83,Munczek91,Wangkl93}. The FBP is a
sum of Yukawa potentials, which not only  satisfies  chiral
invariance and fully relativistic covariance, but also suppresses
the singular point that the
 Yukawa potential has. It works well in
 understanding the dynamical chiral symmetry breaking, confinement and the QCD vacuum as well as
the meson  structures, such as electromagnetic form factors, radius,
decay constants \cite{WYW03,WYW05,WangWan,Wang02}. In this article,
we use the  FBP to study  the ground state mass spectrum of the
scalar diquarks without fine tuning such as modifying  the infrared
behavior  for the heavy quark systems.

The article is arranged as follows:  we introduce the FBP in section
II; in section III, IV and V, we solve the rainbow SDE and ladder
BSE, explore the analyticity of the quark propagators, study the
dynamical symmetry breaking   and confinement, finally obtain the
mass spectrum of the $\pi$, $K$ mesons and the scalar $U^a$, $D^a$,
$S^a$ diquarks, and the decay constants of the $\pi$, $K$ mesons;
section VI is reserved for conclusion.

\section{Flat Bottom Potential }
The present techniques in QCD  calculation can not give satisfactory
large $r$ behavior for the gluon two-point Green's function to
implement the linear potential confinement mechanism, in practical
calculation, the phenomenological effective
 potential models always do the work.
The FBP is a sum of Yukawa potentials which is an analogy to the
exchange of a series of particles and ghosts with different masses
(Euclidean Form),
\begin{equation}
G(k^{2})=\sum_{j=0}^{n}
 \frac{a_{j}}{k^{2}+(N+j \rho)^{2}}  ,
\end{equation}
where $N$ stands for the minimum value of the masses, $\rho$ is their mass
difference, and $a_{j}$ is their relative coupling constant.
 Due to the particular condition we take for the FBP,
there is no divergence in solving the SDE.
In its three dimensional form, the FBP takes the following form:
\begin{equation}
V(r)=-\sum_{j=0}^{n}a_{j}\frac{{\rm e}^{-(N+j \rho)r}}{r}  .
\end{equation}
In order to suppress the singular point at $r=0$, we take the
following conditions:
\begin{eqnarray}
V(0)=constant, \nonumber \\
\frac{dV(0)}{dr}=\frac{d^{2}V(0)}{dr^{2}}=\cdot \cdot
\cdot=\frac{d^{n}V(0)} {dr^{n}}=0    .
\end{eqnarray}
The  $a_{j}$ can be  determined by solving  the equations  inferred
from the flat bottom condition in Eq.(3). As in  previous literature
\cite{WYW03,Wangkl93,WYW05,WangWan,Wang02}, $n$ is set to be 9.

\section{Schwinger-Dyson equation}
The SDE can provide a natural
  framework for studying the nonperturbative properties  of the
  quark and gluon Green's functions. By studying the evolution
  behavior and analytic structure of the dressed quark propagators,
  we can obtain valuable information about the dynamical chiral symmetry breaking and confinement.
 In the following, we write down the rainbow SDE for the quark propagator,
\begin{equation}
S^{-1}(p)=i\gamma \cdot p + \hat{m}_{u,d,s}+ 4\pi \int \frac
{d^{4}k}{(2 \pi)^{4}} \gamma_{\mu}\frac{\lambda^a}{2}
S(k)\gamma_{\nu}\frac{\lambda^a}{2}G_{\mu \nu}(k-p),
\end{equation}
where
\begin{eqnarray}
S^{-1}(p)&=& i A(p^2)\gamma \cdot p+B(p^2)\equiv A(p^2)
[i\gamma \cdot p+m(p^2)], \\
G_{\mu \nu }(k)&=&(\delta_{\mu \nu}-\frac{k_{\mu}k_{\nu}}{k^2})G(k^2),
\end{eqnarray}
and $\hat{m}_{u,d,s}$ stands for the current quark mass that
 breaks chiral symmetry explicitly. For a short discussion about the
full  SDE for the quark propagator, one can consult
Ref.\cite{WYW05}.

In this article, we assume that a Wick rotation to Euclidean
variables is allowed, and perform a rotation analytically continuing
$p$ and $k$ into the Euclidean region.  Alternatively, one can
derive the SDE from the  Euclidean path-integral formulation of the
theory, thus avoiding
 possible difficulties in performing the Wick
 rotation $\cite{Stainsby}$ . As far as only numerical results are concerned,
  the two procedures are equal. In fact, the analytical  structures of quark propagators have
 interesting information about confinement, we will make detailed discussion about
the  propagators of  the $u$, $d$ and $s$ quarks in  section V.

\section{Bethe-Salpeter equation}
The BSE is a conventional approach in dealing with the two-body
relativistic bound state problems \cite{BS51}. The precise knowledge
about the quark structures of the mesons and  diquarks can result in
better understanding of their properties. In the following, we write
down the ladder BSE for the  scalar diquark quasi-bound states  with
two quarks of different flavor \cite{Roberts87,BurdenM},
\begin{eqnarray}
\lefteqn{\Gamma_{\bar 3}(q,P) =  }\\
&& \nonumber - 4\pi\int \frac{d^4k}{(2\pi)^4} G_{\mu\nu}(q - k)
\gamma_\mu \frac{\lambda^a}{2}\,S(k + \xi P)
        \Gamma_{\bar 3}(k,P)  S^T(-k + (1-\xi) P)
    (\gamma_\nu\frac{\lambda^a}{2})^T .
\end{eqnarray}
Here $T$ denotes matrix transpose,  the $S(k)$ is the quark
propagator, $G_{\mu \nu}(k)$ is the gluon propagator, $P_\mu$ is the
four-momentum of the center of mass of the scalar diquark, $q_\mu$
is the relative four-momentum between the two quarks, $\gamma_{\mu}$
is the bare quark-gluon vertex,  and
 $\Gamma_{\bar{3}}(q,P)$ is the Bethe-Salpeter amplitude  of the quasi-bound
 state (or diquark). The $\xi$ is the center of mass parameter which can be chosen to  vary between $0$ and
 $1$, for the $S^a$ diquark, $\xi=\frac{1}{2}$, for the $U^a$ and
 $D^a$ diquarks, as the current quark masses  $m_s>m_u$ and $m_s>m_d$,
 $\xi$ is about $ \frac{1}{2}$.
We can introduce an auxiliary amplitude $\Gamma_{\bar 3}^C(q,P)$ to
facilitate the calculation,
\begin{equation}
\Gamma_{\bar 3}^C(q,P) \equiv \Gamma_{\bar 3}(q,P) \,C,
\end{equation}
here $C=\gamma_2\gamma_4$ is the charge conjugation matrix.  The
auxiliary amplitude $\Gamma_{\bar 3}^C(q,P)$ satisfies the following
equation,
\begin{equation}
\Gamma_{\bar 3}^C(q,P) =  -\frac{8\pi}{3} \int \frac{d^4
k}{(2\pi)^4} G_{\mu\nu}(q - k) \gamma_\mu\,S(k + \xi P)\,
\Gamma_{\bar 3}^C(k,P)
 S(k - (1-\xi) P) \gamma_\nu.
\end{equation}
 It is  obviously the above equation is
identical to the BSE for the pseudoscalar mesons but a reduction in
the  coupling strength, $\frac{4}{ 3}\rightarrow \frac{2}{ 3}$.
 We can introduce the Bethe-Salpeter wavefunction (BSW) $\chi_{qq}$ for
 the quasi-bound states,
 \begin{equation}
\chi_{qq}(q,P)\equiv S(q + \xi P)\Gamma_{\bar 3}^C(p,P)S(q - (1-\xi)
P) ,
\end{equation}
to relate with  our previously works on the pseudoscalar meson's
BSE,
\begin{eqnarray}
S^{-1}(q+\xi P)\chi_{qq}(q,P)S^{-1}(q-(1-\xi) P)=-\frac{8 \pi }{3}
\int \frac{d^4 k}{(2\pi)^4}\gamma_\mu \chi_{qq}(k,P)
\gamma_\nu G_{\mu \nu}(q-k),\\
 S^{-1}(q+\xi
P)\chi_{q\bar{q}}(q,P)S^{-1}(q-(1-\xi) P)=-\frac{16 \pi }{3} \int
\frac{d^4 k}{(2\pi)^4}\gamma_\mu \chi_{q\bar{q}}(k,P) \gamma_\nu
G_{\mu \nu}(q-k),
\end{eqnarray}

We can perform the Wick rotation analytically and continue  $q$
and $k$ into the Euclidean region \footnote{To avoid possible
difficulties in performing the Wick rotation, one can derive the
BSE from the Euclidean path-integral formulation of the theory. }.
 The BSWs of the scalar diquarks $\chi_{qq}$ and pseudoscalar
 mesons $\chi_{q\bar{q}}$ have the same Dirac structures,
 can be signed by the notation $\chi(q,P)$.  In the lowest order approximation, the BSW $\chi(q,P)$ can be
 written as
\begin{eqnarray}
\chi(q,P)=\gamma_5 \left[ iF_1^{0}(q,P)+\gamma \cdot P F_2^{0}(q,P)
+\gamma \cdot q q\cdot P F_3^{1}(q,P)+i[\gamma \cdot q,\gamma \cdot P  ] F_4^{0}(q,P) \right].
\end{eqnarray}
In solving the BSEs, it is important to translate the wavefunctions
$F_{i}^{n}$ into the same dimension,
\begin{eqnarray}
F_{1}^{0}\rightarrow \Lambda^{0}F_{1}^{0}, \, F_{2}^{0}\rightarrow
\Lambda^{1}F_{2}^{0}, \, F_{3}^{1}\rightarrow \Lambda^{3}F_{3}^{1},
\,F_{4}^{0}\rightarrow \Lambda^{2}F_{4}^{0}, \, q\rightarrow
q/\Lambda, \, P\rightarrow P/\Lambda \, , \nonumber
\end{eqnarray}
here the  $\Lambda$ is some quantity of the dimension of mass.
 The ladder BSEs for the scalar diquarks and pseudoscalar mesons  can be projected into the following four coupled integral equations,
\begin{eqnarray}
\sum_j H(i,j)F_j^{0,1}(q,P)&=&\sum_j \int d^4k K(i,j) ,
\end{eqnarray}
the expressions of the $H(i,j)$ and $K(i,j)$ are cumbersome and neglected here.

We can introduce a parameter $\lambda(P^2)$ and solve the above
equations as an eigenvalue problem.  If there really exist a
quasi-bound state of two-quark, the masses of the diquarks can be
determined by the condition $\lambda(P^2=-M_{qq}^2)=1$,
\begin{eqnarray}
\sum_j H(i,j)F_j^{0,1}(q,P)&=&\lambda(P^2)\sum_j \int d^4k K(i,j) .
\end{eqnarray}

Here we will take a short digression and  give some explanations for
the expressions of $H(i,j)$ . The $H(i,j)$'s are functions of the
quark's Schwinger-Dyson functions (SDF) $A(q^2+\xi^2 P^2+ 2\xi q
\cdot P)$ , $B(q^2+\xi^2 P^2+ 2\xi q \cdot P)$, $A(q^2+
(1-\xi)^2P^2-2(1-\xi) q \cdot P)$ and $B(q^2+ (1-\xi)^2P^2-2(1-\xi)
q \cdot P)$ . The relative four-momentum $q$ is a quantity in the
Euclidean spacetime while the center of mass four-momentum $P$ must
be continued to the Minkowski spacetime i.e.
$P^2=-m^2_{\pi,K,U^a,D^a,S^a}$ , this results in that the $q \cdot
P$ varies  throughout a complex domain. It is inconvenient to solve
the SDE with the resulting complex values of the quark momentum. We
can expand the $A$ and $B$ in terms of Taylor series of  $q \cdot
P$, for example,
\begin{eqnarray}
A(q^2+\xi^2P^2+ \xi q \cdot P)&=&A(q^2+\xi^2P^2)+2\xi A(q^2+\xi^2
P^2)' q \cdot P+\cdots. \nonumber
 \end{eqnarray}
The other problem is that we can not solve the SDE in the timelike
region as the two-point gluon Green's function can not be exactly
inferred from the $SU(3)$ color gauge theory even in the low energy
spacelike region. In practical calculations, we can extrapolate the
values of the  $A$ and $B$ from the spacelike region smoothly to the
timelike region with suitable  polynomial functions. To avoid
possible violation with confinement in sense of the appearance of
pole masses $q^2=-m^2(q^2)$ in the timelike region, we must be care
in
  choosing the polynomial functions \cite{Munczek91}.

 Finally we write down the normalization condition for
the BSWs of the pseudoscalar mesons,
\begin{eqnarray}
N_c \int \frac{d^4q}{(2\pi)^4} Tr \left\{ \bar{\chi}
\frac{\partial S^{-1}_{+}} {\partial P_{\mu}}\chi(q,P) S^{-1}_{-}
+\bar{\chi} S^{-1}_{+} \chi(q,P) \frac{\partial S^{-1}_{-}}
{\partial P_{\mu}} \right\}=2 P_{\mu},
\end{eqnarray}
here $\bar{\chi}=\gamma_4 \chi^+ \gamma_4$, $S_+=S(q+\xi P)$ and
$S_-=S(q-(1-\xi)P)$. In this article, the parameters of FBP are
fitted to give the correct masses and decay constants for the
pseudoscalar mesons, $\pi$ and $K$, the normalization condition is
needed.

\section{Coupled rainbow SDE and ladder BSE, and the mass spectrum}
In this  section, we study  the coupled equations of the rainbow SDE
and ladder BSE for the pseudoscalar mesons ($\pi$ and $K$) and
scalar diquarks ($U^a$, $D^a$ and $S^a$) numerically, the final
results for the SDFs and BSWs can be plotted as functions of the
square momentum $q^2$.

In order to demonstrate the confinement of quarks, we have to study
the analyticity of the SDFs of the  $u$, $d$ and $s$ quarks, and
prove that there are no mass poles on the real timelike  $q^2$
axial.  In the following, we  take the Fourier transform
 with respect to the Euclidean time T
 for the scalar part ($S_{s}$) of the quark propagator \cite{Roberts94,Roberts00,Maris95},
 \begin{eqnarray}
 S^{*}_{s}(T)  =  \int_{-\infty}^{+ \infty} \frac{dq_{4}}{2 \pi} e^{iq_{4}T}
 \frac{B(q^2)}{q^2A^2(q^2)+B^{2}(q^2)}|_{ \overrightarrow{q}=0},
 \end{eqnarray}
where the 3-vector part of $q$ is set to zero.
 If  S(q) has a mass pole at $q^2=-m^2(q^2)$ in the real timelike region, the Fourier transformed
  $S^{*}_{s}(T)$ would fall off as $e^{-mT}$ for large T or
  $\log{S^{*}_{s}}=-mT$.
In our numerical calculations, for small $T$, the values of
$S^{*}_{s}$ are positive  and  decrease rapidly to zero and beyond
with the increase  of $T$, which are compatible with the result
(curve tendency with respect to $T$) from  lattice simulations
\cite{Bhagwat03} ; for large $T$, the values of $S^{*}_{s}$ are
negative, except occasionally a very small fraction of positive
values.  The negative values for $S^{*}_{s}$ indicate  an explicit
violation of the axiom of reflection positivity \cite{Jaffee},
 in other words, the quarks are not physical observable i.e.
 confinement. As colored quantity, the diquarks should also be confined and not appear
 as asymptotic states; the demonstration of their confinement is beyond the present work.

The $u$, $d$ and $s$ quarks have small current masses, the dressing
or renormalization is very
 large and the curves of the SDFs are  steep, which are  corresponding  to the dynamical chiral
symmetry breaking phenomenon for the light quarks.
 At zero momentum, $m_u(0)=m_d(0)=0.454 GeV $ and $m_s(0)=0.684 GeV $,  which
  are compatible with the constituent quark masses
in the literature. From the solutions of BSEs for the $\pi$, $K$
mesons and $U^a$, $D^a$, $S^a$ diquarks as eigenvalue problems, we
can obtain the masses for those pseudoscalar mesons and scalar
diquarks,
\begin{eqnarray}
M_{\pi}=135MeV, \,\,\,  M_{K}=498MeV, \nonumber \\
M_{S^a}=0.76GeV, \,\,\,M_{U^a}=0.98GeV, \,\,\,M_{D^a}=0.98MeV.
\end{eqnarray}
It is obviously
\begin{eqnarray}
M_{S^a}<m_u(0)+m_d(0), \,\,\,M_{U^a}<m_d(0)+m_s(0),
\,\,\,M_{D^a}<m_u(0)+m_s(0), \\
M_{U^a}-M_{S^a}\approx m_s(0)-m_u(0)\approx 0.22GeV.
\end{eqnarray}
The attractive interaction between the quarks in the color and
flavor $SU(3)$ $\bar{3}$ channel can lead to the quasi-bound states
in the infrared region. The appearance of the diquarks is closely
related to the dynamical symmetry breaking phenomenon, the mass
splitting among the $U^a$, $D^a$ and $S^a$ diquarks originate from
the mass splitting among $u$, $d$ and $s$ quarks.

The existing theoretical calculations for the masses of the scalar
diquarks vary in a large range, $M_{qq}=(0.4-0.7)GeV $
 with QCD sum rules \cite{DoschM}; $M_{qq}\sim 0.5 GeV$ with random
 instanton liquid model \cite{RipkaM}; $M_{qq}= (0.42\pm 0.03) GeV$ with random
 instanton liquid model \cite{ShuryakM},
 $M_{qq}=0.234GeV$ with Nambu-Jona-Lasinio Model \cite{VoglM}; $M_{S^a}=0.74GeV,
 M_{U^a}=M_{D^a}=0.88GeV$ with BSE \cite{BurdenM}; $M_{S^a}=0.82GeV,
 M_{U^a}=M_{D^a}=1.10GeV$ with BSE \cite{MarisM}; $M_{qq}= 0.692 GeV$
 with global color model \cite{CahillM}; $0.14<M_{qq}< 0.74 GeV$
 with assumption of mass functions  \cite{Roberts87};  $M_{qq}\approx 0.7 GeV$ with lattice QCD \cite{HessM}.
 There are large uncertainties  for the masses of the quasi-bound
 states, $U^a$, $D^a$, and $S^a$. As colored quantities, the diquarks may have gauge interactions
 with the gluon field as fundamental scalar field with the following
 Lagrangian,
 \begin{eqnarray}
L&=&-\frac{1}{2}D_\mu {S^a}^+ D_\mu
S^a-\frac{1}{2}M_{S^a}^2{S^a}^+S^a , \\
D_\mu S^a&=&\partial_\mu S^a+ig f^{abc} A^b_\mu S^c.
 \end{eqnarray}
Here we use  the notation $S^a$ to represent the scalar diquark
field and $A^b_\mu$ the gluon field. The direct color interactions
between the scalar diquarks and the gluons may  modify the mass
$M_{qq}$ significantly.

\begin{figure}
 \centering
 \includegraphics[totalheight=7cm]{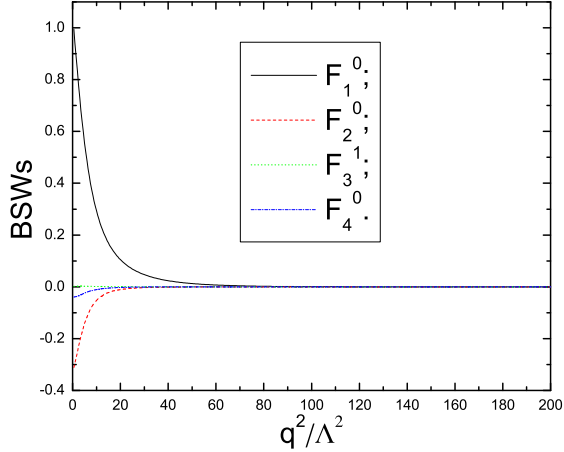}
 \caption{Un-normalized BSWs for the $\pi$ meson. }
\end{figure}
\begin{figure}
 \centering
 \includegraphics[totalheight=7cm]{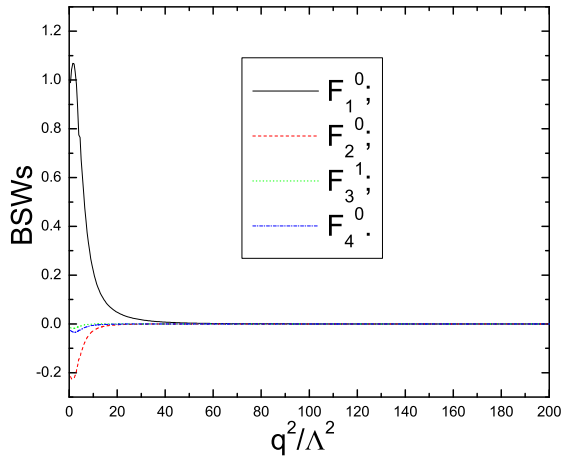}
 \caption{Un-normalized BSWs for the $S^a$ diquark .}
\end{figure}
 From the plotted BSWs (see Fig.1 for the
$\pi$ meson and Fig.2 for the $S^a$ diquark as examples), we can see
that the BSWs of the pseudoscalar mesons and scalar diquarks  have
the same type (Gaussian type) momentum dependence while
   the quantitative values are different from each other. Just like the $\bar{q}q$ ,
   $\bar{q}Q$ and $\bar{Q}Q$
   pseudoscalar  mesons \cite{WYW03,WYW05},
   the gaussian type BSWs of the scalar diquarks
center around  zero momentum and  extend to the energy scale about
$q^2=1GeV^2$ which happen to be the energy scale for the chiral
symmetry breaking, the strong interactions in the infrared region
result  in quasi-bound states, $U^a$, $D^a$ and $S^a$. The BSWs of
the $\pi$ and $K$ mesons can give satisfactory  values for the decay
constants which are defined by
\begin{eqnarray}
i f_{\pi} P_\mu &=& \langle0|\bar{q}\gamma_\mu \gamma_5 q |\pi(P)\rangle, \nonumber \\
&=& N_c \int Tr \left[\gamma_\mu \gamma_5\chi(k,P)\right]
\frac{d^4 k}{(2\pi)^4} ,
\end{eqnarray}
here we use $\pi$ to represent the pseudoscalar mesons,
\begin{eqnarray}
f_{\pi}=127 MeV;  \, \, \, f_{K}=161 MeV.
\end{eqnarray}

There are  negative voice  for the existence of the scalar diquark
states  in the infrared region \cite{NoDiquark}. The coupled rainbow
SDE and ladder BSE are particularly suitable for studying the flavor
octet pseudoscalar mesons and vector mesons, the
next-to-leading-order (NLO) contributions from the  quark-gluon
vertex have a significant amount of cancellation between repulsive
and attractive corrections. However,  for the diquarks and scalar
mesons, the large repulsive corrections from the NLO contributions
can significantly change the scalar meson masses  and corresponding
BSWs; the diquark quasi-bound states found in the ladder BSE will
disappear from the spectrum. In this article, we do not take the
point of view that the scalar diquarks exist in the strong
interaction spectrum as asymptotic states, and our conclusion do not
conflict with Ref.\cite{NoDiquark}; the confinement precludes the
observation of the free colored diquarks, the quark-quark can
correlate with each other in the color and flavor $\bar{3}$ channels
inside the baryons and multi-quark states with typical length
$l=\frac{1}{M_{qq}}$. In fact, the NLO contributions from the
quark-gluon vertex are extremely  difficult to take into account if
we go beyond the infrared dominated $\delta$ function approximation
for the gluon kernel, which is obviously violate the lorentz
invariance; the lorentz invariant  and model-independent treatments
still lack in the literatures.

 In calculation, the values of current quark masses are taken
as $\hat{m}_u=\hat{m}_d=6MeV$ and $\hat{m}_s=150 MeV$; the input
parameters for the FBP are $N=1.0 \Lambda $, $V(0)=-17.0 \Lambda$,
 $\rho=6.0\Lambda$ and $\Lambda=200 MeV$, which are fitted to give the correct masses of the $\pi$ and $K$ mesons.
   In this article,  we deal with only  the light flavor quarks, the FBP can give
 satisfactory results without fine tuning, such as modifying the
 infrared behavior   for the heavy quarks   $c$ and $b$.

\section{Conclusion }
In this article, we study the under-structures of the pseudoscalar
mesons $\pi$ , $K$ and scalar diquarks $U^a$, $D^a$, $S^a$ in the
framework of the coupled rainbow SDE and ladder BSE with the
confining effective potential (FBP). After we solve the coupled
rainbow SDE and ladder BSE numerically, we obtain the SDFs and BSWs
of the pseudoscalar mesons $\pi$ , $K$ and scalar diquarks $U^a$,
$D^a$, $S^a$, and the corresponding ground state mass spectrum. The
$u$, $d$ and $s$ quarks have small current masses,  the dressing or
renormalization for the SDFs is very large and the curves are  steep
which indicate the  dynamical chiral symmetry breaking phenomenon
for the light quarks explicitly.  The mass poles in the timelike
region are absent which  implement the confinement naturally. The
BSWs of the pseudoscalar mesons and scalar diquarks have the same
type (Gaussian type) momentum dependence while the quantitative
values are different from each other. The gaussian type BSWs
    center around  zero momentum and extend to the energy scale about
$q^2=1GeV^2$ which happen to be the energy scale for the chiral
symmetry breaking, the strong interactions in the infrared region
result  in bound (or quasi-bound) states. Our numerical results for
the masses and decay constants of the $\pi$, $K$ mesons can
 reproduce the experimental values, the ground state
mass spectrum of scalar diquarks are consistent with the existing
theoretical calculations. The mass splitting among  the $U^a$, $D^a$
and $S^a$ diquarks originate from the mass splitting among $u$, $d$
and $s$ quarks. $M_{S^a}<m_u(0)+m_d(0)$, $M_{U^a}<m_d(0)+m_s(0)$ ,
$M_{D^a}<m_u(0)+m_s(0)$ , the attractive interaction between the
color and flavor $SU(3)$ $\bar{3}$ quarks can lead to the
quasi-bound states in the infrared region.   Once the satisfactory
SDFs and BSWs of  the scalar diquarks are known, we can use them to
study a lot of important quantities involving the multiquark states.

\section*{Acknowledgment}
This  work is supported by National Natural Science Foundation,
Grant Number 10405009,  and Key Program Foundation of NCEPU. The
authors are indebted to Dr. J.He (IHEP), Dr. X.B.Huang (PKU) and Dr.
L.Li (GSCAS) for numerous help, without them, the work would not be
finished. The authors would also like to thank Prof. C. D. Roberts
for providing us some important  literatures.

\end{document}